\documentclass[aps,pre,preprint,superscriptaddress,showpacs,showkeys]{revtex4}

\usepackage{graphicx}% Include figure files
\usepackage{dcolumn}% Align table columns on decimal point
\usepackage{bm}% bold math
\usepackage{amssymb}
\usepackage{amsmath}

\usepackage{color}

\newcommand{\alkor}[1]{\textcolor{red}{#1}}
\renewcommand{\alkor}[1]{#1}
\newcommand{\aeh}[1]{\textcolor{red}{#1}}
\renewcommand{\aeh}[1]{#1}

\begin{document}

\title{\alkor{Incomplete} noise-induced synchronization of spatially extended systems}
\author{Alexander~E.~Hramov}
%\email{aeh@nonlin.sgu.ru}
\author{Alexey~A.~Koronovskii}
%\email{alkor@nonlin.sgu.ru}
\author{Pavel V. Popov}
%\email{popovpv@nonlin.sgu.ru}
 \affiliation{Faculty of Nonlinear
Processes, Saratov State University, Astrakhanskaya, 83, Saratov,
410012, Russia}

\begin{abstract}
A new type of noise-induced synchronous behavior is described. This
phenomenon, called \textit{\alkor{incomplete} noise-induced
synchronization}, arises for one-dimensional Ginzburg-Landau
equations driven by common noise. The mechanisms resulting in the
\alkor{incomplete} noise-induced synchronization in the spatially
extended systems are revealed analytically. The different model
noise are considered. A very good agreement between the theoretical
results and the numerically calculated data is shown.
\end{abstract}

\date{\today}

\pacs{05.45.-a, 05.40.-a, 05.45.Xt}

\keywords{spatially extended systems, \alkor{incomplete}
noise-induced synchronization, modified system approach, complex
Ginzburg-Landau equation}

\noindent{\sf{Hramov A.E., Koronovskii A.A., Popov P.V. Incomplete
noise-induced synchronization of spatially extended systems. Phys.
Rev. E. 77, (2008) 036215}}

\vspace {20mm}

\maketitle

\section*{Introduction}

Noise-induced synchonization~\cite{Fahy:1992_NoiseInfluence,%
Martian:1994_SynchroNoise,Herzel:1995_NoiseSynchroReconsidered} is
an ubiquitous phenomenon in nonlinear science. It arises as the
interplay between determined and random
dynamics~\cite{Hramov:2006_PLA_NIS_GS}, with both the
synchronization and noise influence being recently the subjects of
considerable interest of scientific community. Indeed, on the one
hand the synchronous behavior of nonlinear systems has attracted
great attention of researchers for a long
time~\cite{Pecora:1990_ChaosSynchro,Rosenblum:1996_PhaseSynchro,%
Boccaletti:2002_ChaosSynchro,Hramov:2004_Chaos}. On the other hand
discovering the fact that fluctuations can actually induce some
degree of order in a large variety of nonlinear systems is one of
the most surprising results of the last decades in the field of
stochastic processes~\cite{Pikovsky:1997_CoherenceResonance,%
Mangioni:1997_Noise,Zaikin:2000_DblStochRes}. Moreover, both these
phenomena are relevant for physical, chemical, biological and other
systems described in terms of nonlinear dynamics (see, e.g.,~\cite{Shuai:1998_NoiseSynchro,%
Neiman:2002_SynchroAndNoise,Zhou:2002_NoiseEnhancedPS,Zhou:PRE2003}).

Noise-induced synchronization (NIS) means that the random signal
influencing two identical uncoupled dynamical chaotic systems
$\mathbf{u}(t)$ and $\mathbf{v}(t)$ (starting from the different
initial conditions $\mathbf{u}(t_0)$ and $\mathbf{v}(t_0)$,
$\mathbf{u}(t_0)\not=\mathbf{v}(t_0)$) results in their synchronous
(i.e., identical) behavior $\mathbf{u}(t)=\mathbf{v}(t)$ after
transient finished.

Noise-induced synchronization can be detected by means of direct
comparison of the states of two chaotic systems, $\mathbf{u}(t)$,
and, $\mathbf{v}(t)$, being under the influence of noise. The
other method of diagnostics of NIS is calculating the largest
Lyapounov exponent (LE) of dynamical system that measures the
stability of the motion. Indeed, in driven chaotic system the
largest Lyapunov exponent may become negative, that results in
synchronization: both systems forced by the same noise ``forget''
their initial conditions and evolve to identical
state~\cite{Goldobin:2005_SynchroCommonNoise}. If the noise
influence is \aeh{infinitely} small the largest Lyapunov exponent
is positive for such a system.

In all cases of the noise-induced synchronization being known
hitherto the boundary of the noise-induced synchronization regime
is associated with the point on the parameter axis where the
largest Lyapunov exponent of the system under study crosses the
zero value when its sign is changed from ``plus'' to ``minus''. In
this paper we report for the first time that the noise-induced
synchronization regime of two spatially extended uncoupled
identical systems driven by common noise may be preceded by a new
type of behavior, \aeh{when the largest Lyapunov exponent remains
zero in a finite range of parameter values}. This kind of behavior
called \emph{``\alkor{incomplete} noise induced synchronization''}
(INIS) demonstrates the features of the synchronous motion of two
uncoupled identical systems driven by common noise: \aeh{although
the states of the system differ, trajectories can be transformed
into each other by an appropriate spatial shift}.

The structure of the paper is the following. In Sec.~\ref{sct:GLEs}
we describe the system under study being the complex
Ginzburg--Landau equations driven by common noise and the new type
of their behavior (\alkor{incomplete} noise induced synchronization)
observed for the certain set of the values of the control
parameters. In Sec.~\ref{sct:Mechanisms} we consider the mechanisms
being responsible for the INIS regime.
Sec.~\ref{sct:NoiseCharacteristics} deals with the different models
of noise with the distinct probability densities. The final
conclusions are given in Sec.~\ref{sct:Conclusion}.

\section{System under study and incomplete noise-induced synchronization}
\label{sct:GLEs}

The system under study is represented by a pair of uncoupled
complex Ginzburg--Landau equations (CGLEs) driven by common noise:
\begin{equation}
\begin{array}{l}
\displaystyle
 u_t= u-(1-i\beta)|u|^2
 u+(1+i\alpha) u_{xx} + D\zeta(x,t), \\
\displaystyle
 v_t= v-(1-i\beta)|v|^2
 v+(1+i\alpha) v_{xx} + D\zeta(x,t), \\
\end{array}
 \label{eq:G-L_noise}
\end{equation}
where $u(x,t)$, $v(x,t)$ are complex states of the considered
systems, $\alpha$ and $\beta=4$ are the control parameters, $D$
defines the intensity of a noise term
\aeh{$\zeta=\zeta_1+i\zeta_2$}. We have used model noise with the
asymmetrical probability distribution of the real $\aeh{\zeta_1}$
and \aeh{imaginary} $\aeh{\zeta_2}$ parts of the random variable
\begin{equation}
p(\aeh{\zeta_{1,2}})=\left\{
\begin{array}{ll}
2\aeh{\zeta_{1,2}}, & \mathrm{if~0\leq\aeh{\zeta_{1,2}}\leq1,}\\
0, & \mathrm{otherwise}\\
\end{array}\right.
\label{eq:NoiseDistribution}
\end{equation}
on the unit interval $\aeh{\zeta_{1,2}\in}[0;1]$. The simulation
of the random variables $\zeta\aeh{_{1,2}}$ with required
probability distribution $p(\zeta\aeh{_{1,2}})$ was carried out in
the same way as it was described in~\cite{Sweet:2001_Step&Stagger}
for the exponential stagger distribution. Equation
(\ref{eq:G-L_noise}) was solved with periodic boundary conditions
\begin{equation}
{u(x,t)=u(x+L,t)} \mbox{\quad and\quad} {v(x,t)=v(x+L,t)},
\label{eq:PeriodicBoundaryConditions}
\end{equation}
with all numerical calculations being performed for a fixed system
length $L=40\pi$ and random initial conditions. To evaluate
(\ref{eq:G-L_noise}) the standard numerical scheme has been
used~\cite{Garcia-Ojalvo:1999_NoiseBook}, the value of the grid
spacing is $\Delta x=L/1024$, the time step of the scheme $\Delta
t=2.0\times 10^{-4}$.

\begin{figure}[tb]
%\vspace*{4cm}
\begin{center}
\includegraphics*[scale=0.35]{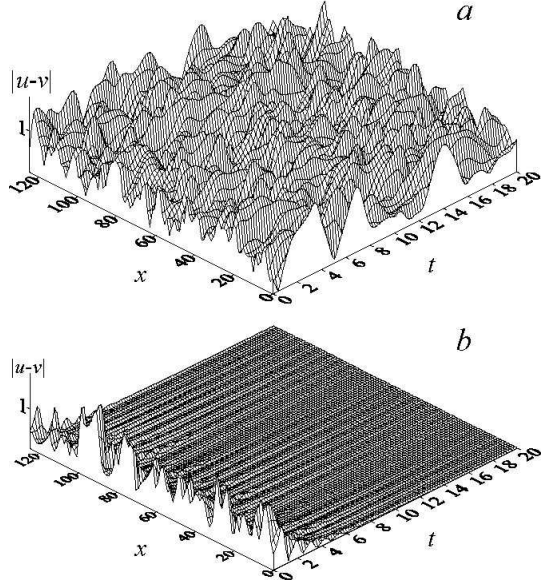}
\end{center}
\caption{The evolution of the difference of the system states
${|u(x,t)-v(x,t)|}$ described by complex Ginzburg-Landau
equations~(\ref{eq:G-L_noise}) (\textit{a}) without noise and
(\textit{b}) with noise with the intensity $D=3$.
\aeh{In the second case the difference of the states of both systems
in every point of space tends to be zero (after transient), which
means the presence of the noise-induced synchronization regime.}
The control parameter values are $\alpha=2$, $\beta=4$
\label{fgr:Differences}}
\end{figure}

If the noise intensity is equal to zero ($D=0$) and initial
conditions $u(x,0)$ and $v(x,0)$ are not identical, both systems
demonstrate the complex chaotic behavior (both in time and in
space), with the system states being different, i.e., ${u(x,t)\neq
v(x,t)}$ (Fig.~\ref{fgr:Differences},\textit{a}). Alternatively, if
the noise intensity $D$ is large enough the states of both systems
coincide with each other (Fig.~\ref{fgr:Differences},\textit{b}),
that is the evidence of the noise-induced synchronization.

To detect the presence of the noise--induced synchronization regime
the averaged difference
\begin{equation}
\varepsilon=\frac{1}{TL}\int\limits_\tau^{\tau+T}\int\limits_0^L|u(x,t)-v(x,t)|\,dxdt,
\label{eq:epsilon}
\end{equation}
between the spatio-temporal states of two CGLEs driven by common
noise was calculated. The averaging process starts after a long-time
transient with duration $\tau=200$.

In the NIS regime the relation $\varepsilon=0$ takes place, since
in this case the difference between the states of two identical
spatially extended systems (\ref{eq:G-L_noise}) in every point of
space tends to zero. We have also calculated the largest Lyapunov
exponent $\lambda$ for one of the systems~(\ref{eq:G-L_noise}). As
it was mentioned above in the NIS regime the largest Lyapunov
exponent $\lambda$ should be negative.

\begin{figure}[tb]
%\vspace*{10cm}
\begin{center}
\includegraphics*[scale=0.3]{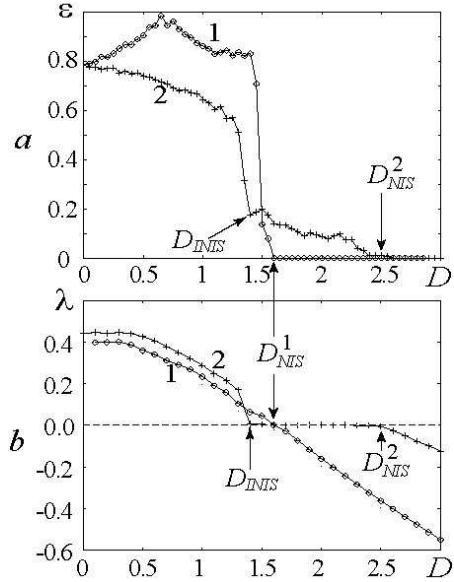}
\end{center}
\caption{The dependencies of (\textit{a}) the averaged difference
(\ref{eq:epsilon}) and (\textit{b}) the largest Lyapounov exponent
of the CGLE on the noise intensity $D$ for the different values of
the control parameter $\alpha$. Curves~1 correspond to the case of
$\alpha=1$, curves~2 were calculated for $\alpha=2$. The values of
noise intensity corresponding to the onset of noise--induced
synchronization are shown by arrows with labels $D^{1}_{NIS}$ and
$D^{2}_{NIS}$ for the curves~1 and 2, respectively. The boundary of
the \alkor{incomplete} noise-induced synchronization is also shown
by arrow marked as $D_{INIS}$ \label{fgr:Diff_and_CLE_on_D}}
\end{figure}

The dependencies of the largest Lyapunov exponent $\lambda(D)$ and
the averaged difference $\varepsilon(D)$ on the noise intensity
$D$ are shown in Fig.~\ref{fgr:Diff_and_CLE_on_D} for two
different values of the control parameter $\alpha$. It is easy to
see that for the control parameter $\alpha=1$ (curves~1 in
Fig.~\ref{fgr:Diff_and_CLE_on_D},\textit{a,b}) the value of the
noise intensity $D$ for which the largest Lyapunov exponent
$\lambda$ crosses zero value and becomes negative coincides with
the point where the averaged difference~(\ref{eq:epsilon}) starts
being vanishingly small. So, in this case the noise--induced
synchronization boundary is $D_{NIS}\approx 1.5$ (see arrows in
Fig.~\ref{fgr:Diff_and_CLE_on_D},\,\textit{a,b}) and we deal with
the occurrence of the noise-induced synchronization regime being
typical and well-known.

Alternatively, a different scenario is observed in the same
system~(\ref{eq:G-L_noise}) if the control parameter value
$\alpha=2$ is considered (see curves~2 in
Fig.~\ref{fgr:Diff_and_CLE_on_D},\,\textit{a,b}). For such a
choice of $\alpha$-parameter value the largest Lyapunov exponent
becomes equal to zero for the large enough intensity of noise
$D_{INIS} \approx 1.53$ whereas the averaged difference
$\varepsilon$ between the spatio-temporal states of two CGLEs
driven by common noise \aeh{exceeds zero value sufficiently}
(Fig.~\ref{fgr:Diff_and_CLE_on_D},\,\textit{a,b}). With further
increase of the noise intensity $D$ (when $D$ is equal to
$D_{NIS}\approx 2.5$) the value of $\varepsilon$ becomes equal to
zero (see Fig.~\ref{fgr:Diff_and_CLE_on_D},\,\textit{a}) and the
largest Lyapunov exponent starts to be negative \aeh{which} is the
evidence of the presence of the noise-induced synchronization
regime.

\begin{figure}[tb]
%\vspace*{4cm}
\begin{center}
\includegraphics*[scale=0.3]{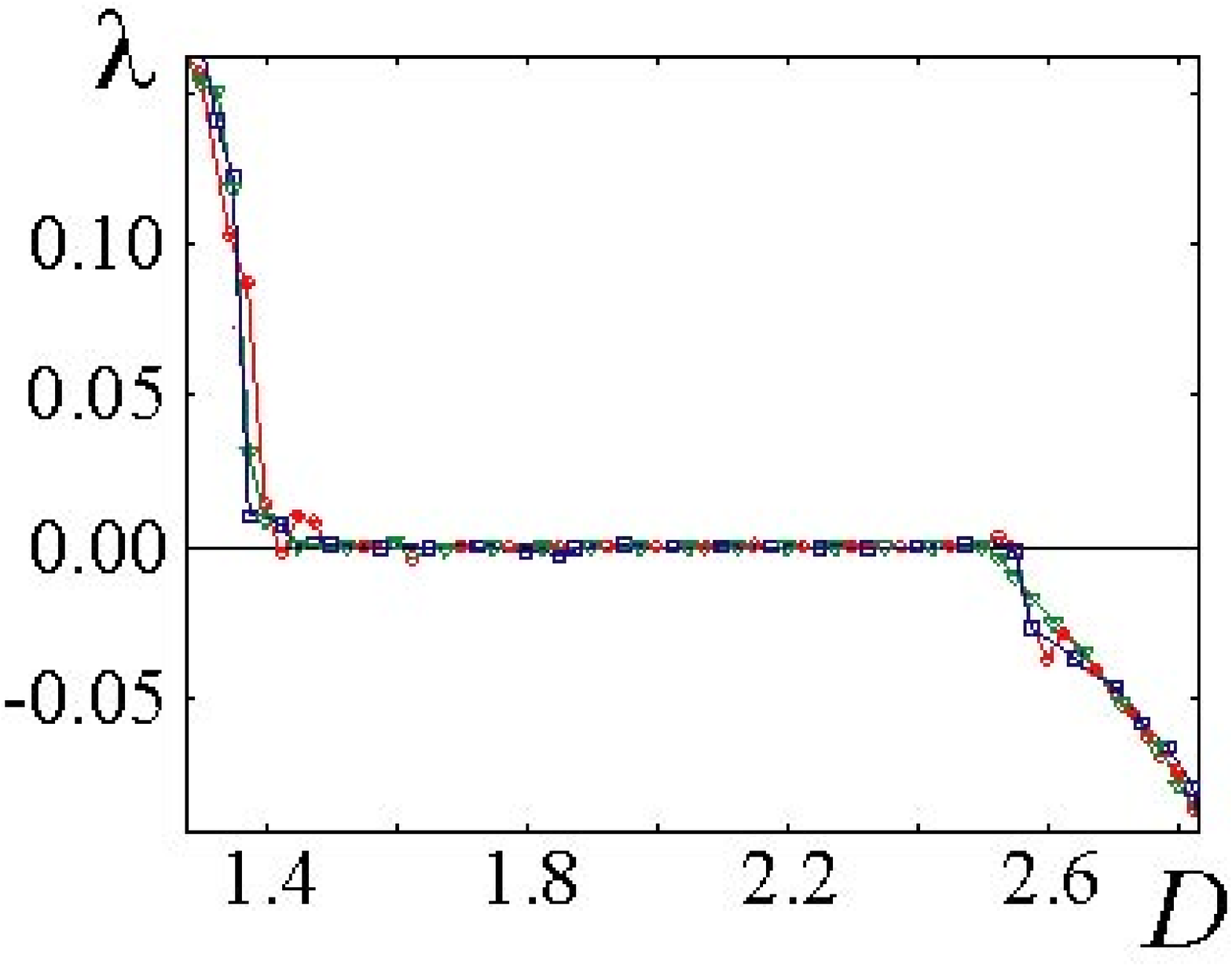}
\end{center}
\caption{(Color online) The dependencies of the largest Lyapunov
exponent $\lambda$ on the noise intensity $D$ calculated for the
different values of the spatial grid spacing:
$\textcolor{blue}{\square}$ --- $\Delta x=L/2^{10}$, $\Delta
t=2.0\times 10^{-4}$, $\textcolor{red}{\lozenge}$
--- $\Delta x=L/2^{11}$, $\Delta t=1.0\times 10^{-4}$, $\textcolor{green}{\triangledown}$ --- $\Delta
x=L/2^{12}$, $\Delta t=5.0\times 10^{-5}$. The plateau where
$\lambda=0$ is shown. The control parameter values are $\alpha=2$,
$\beta=4$ \label{fgr:Plateau}}
\end{figure}

In other words, there is the finite interval of the noise intensity
values $(D_{INIS};D_{NIS})$ for which the noise-induced
synchronization is not observed, and the largest Lyapunov exponent
$\lambda$ is equal to zero. To prove this fact we have calculated
the largest Lyapunov exponent of the complex Ginzburg-Landau
equation for different values of the spatial grid spacing. The range
of the noise intensities corresponding to the plateau where
$\lambda=0$ is shown in Fig.~\ref{fgr:Plateau}. One can see that the
largest Lyapunov exponent calculations with the different values of
the spatial grid step give the similar results. Based on these
calculations we come to conclusion that the largest Lyapunov
exponent is actually equal to zero in the finite range of the noise
intensity.

Despite the fact that the noise-induced synchronization is not
observed in the region where $\lambda=0$ (see
Fig.~\ref{fgr:Diff_and_CLE_on_D}), this range of the noise
intensities corresponds to the behavior showing the features of
synchronous dynamics. The manifestation of synchronism \aeh{can}
be observed if one of the complex media described by the
Ginzburg-Landau equation starts to shift along the second one with
the spatial shift $\delta$. In other words, if one uses the
shifted state of one of the system ${v=v(x+\delta,t)}$ in
Eq.~(\ref{eq:G-L_noise}) the averaged difference $\varepsilon$
changes depending on this shift $\delta$. This movement of one of
the systems supposed to be very slow for the transient to be
completed. \aeh{In this case such a spatial shift $\delta_0$ may
be found that both Ginzburg-Landau equations start to behave
identically, with the largest Lyapunov exponent being equal to
zero.} Therefore, we have called this regime
\emph{``\alkor{incomplete} noise-induced synchronization''}
(INIS).

\begin{figure}[tb]
\centerline{\includegraphics*[scale=0.35]{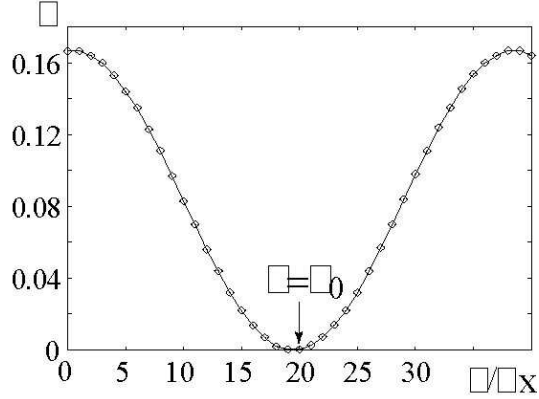}} \caption{The
dependence of the difference $\varepsilon$ between the states of the
media $u(x,t)$ and $v(x,t)$ described by the complex Ginzburg-Landau
equations~(\ref{eq:G-L_noise}) on the space shift $\delta$ for the
control parameters $\alpha=2, \beta=4, D=2$}
\label{fgr:Varepsilon_Delta}
\end{figure}

This statement is illustrated in Fig.~\ref{fgr:Varepsilon_Delta}
where the dependence of the difference $\varepsilon$
(\ref{eq:epsilon}) on the space shift $\delta$ is shown. One can
see that \aeh{there is value $\delta_0$ of shift $\delta$ } for
which the averaged difference $\varepsilon$ becomes equal to zero.
Therefore, for this space shift $\delta_0$ both systems
demonstrate identical behavior and the noise-induced
synchronization is observed. This shift $\delta_0$ depends on the
initial conditions. For the other values of the spatial shift
$\delta$ the system states (both in space and time) are different,
but the largest Lyapunov exponent is always equal to zero for the
considered set of the control parameter values.

\section{Mechanisms resulting in the incomplete noise-induced synchronization regime}
\label{sct:Mechanisms}

Let us discuss the mechanisms resulting in the occurrence of the
\alkor{incomplete} noise-induced synchronization regime. In
work~\cite{Hramov:2006_PLA_NIS_GS} it has been shown, that for
dynamical systems with small number of degrees of freedom the
mechanisms of \aeh{the onset} of noise-induced synchronization and
generalized synchronization are equivalent. The mechanism of the
generalized synchronization occurrence can be considered with the
help of the modified system approach as it was done in
Ref.~\cite{Aeh:2005_GS:ModifiedSystem} for the chaotical systems
with the small number of degrees of freedom and in
Ref.~\cite{Hramov:2005_GLEsPRE} for the spatially extended system.
\aeh{We interpret the mechanism of the onset of incomplete
noise-induced synchronization in a similar way.}
%
%It is possible to assume, that the mechanism of the \aeh{onset of incomplete
%noise-induced synchronization} may be also explained in the same way.
%
Therefore, following Ref.~\cite{Hramov:2006_PLA_NIS_GS,%
Aeh:2005_GS:ModifiedSystem,Hramov:2005_GLEsPRE} we consider the
dynamics of the modified spatially extended system with the
additional term determined by the mean value of noise.

\aeh{The deterministic modified Ginzburg-Landau equation with the
additional term, determined by the mean value $\langle
D\zeta\rangle$ of the noise term $\zeta$ in stochastic
equation~(\ref{eq:G-L_noise})} can be written as
\begin{equation}
 \frac{\partial u_m}{\partial t}= u_m-(1-i\beta)|u_m|^2
 u_m+(1+i\alpha) \frac{\partial^2 u_m}{\partial x^2} + \langle D\zeta\rangle.
 \label{eq:G-L_modified}
\end{equation}
For the selected kind of noise with the probability
distribution~(\ref{eq:NoiseDistribution}) $\langle
D\zeta\rangle=2D/3$.

\alkor{Equation~(\ref{eq:G-L_modified}) \aeh{is forced CGLE},
widely studied and well documented in the literature (see,
e.g.~\cite{Coullet:1992_Patterns_PhysD,Glendinning:1993_GLE_IJBC,Chate:1999_GLE_PhysD}).
It is well-known, \aeh{that different types} of the
spatio-temporal patterns may be observed depending on the domain
of the control parameter values.} If the value $D$ is large
enough, the homogeneous stationary state
${u_0=u_0(x,t)=\mathrm{const}}$ is observed in the
system~(\ref{eq:G-L_modified}). In this case the largest Lyapunov
exponent is negative, with the stationary state regime in the
system~(\ref{eq:G-L_modified}) corresponding to the noise-induced
synchronization in the system~(\ref{eq:G-L_noise}). With decrease
of the noise intensity $D$ the stationary state $u_0$ loses its
stability that corresponds to the boundary of the noise-induced
synchronization of the initial Ginsburg-Landau
equations~(\ref{eq:G-L_noise}) driven by noise.

\begin{figure}[tb]
%\vspace*{10cm}
\begin{center}
\includegraphics*[scale=0.35]{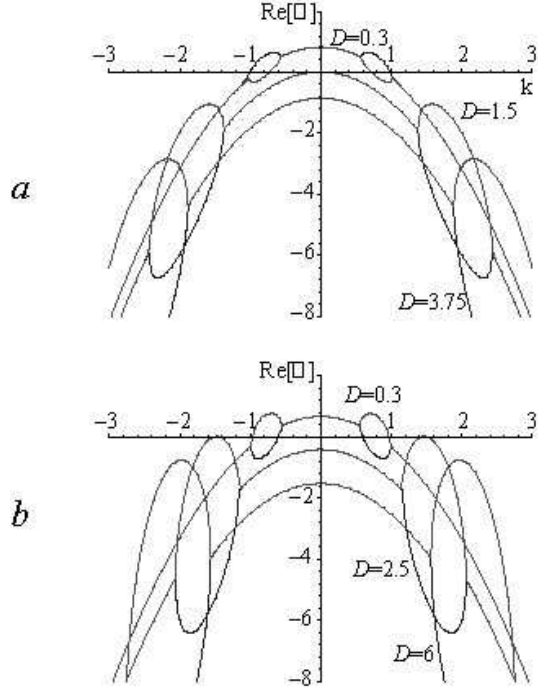}\\
\end{center}
\caption{The dependencies of the real part of the eigenvalues
$\Lambda$ on the wave number $k$ for the different values of
$D$-parameter when the control parameter $\alpha$ has been fixed as
(\textit{a}) $\alpha=1$ and (\textit{b}) $\alpha=2$
\label{fgr:DispRelation}}
\end{figure}

At the same time the loss of the stability of the homogeneous
stationary state occurs \aeh{in  different ways} depending on the
control parameter values of the modified Ginzburg-Landau
equation~(\ref{eq:G-L_modified}).

Indeed, the homogeneous stationary state $u_0$ can be obtained
numerically from equation
\begin{equation}
u_0-(1-i\beta)|u_0|^2u_0+2D/3=0, \label{Eq:ZO}
\end{equation}
e.g., by a Newton method~\cite{NumericalRecipes:1997}. To analyze
the stability of Eq.~(\ref{Eq:ZO}) we have to consider the
linearization of the modified Ginzburg-Landau equation in the
vicinity of the stationary solution $u_0$. Let $\tilde u=\tilde
u_{r}+i\tilde u_{i}$ be a small perturbation of the homogenous
stationary state $u_0=u_r+iu_i$, i.e., $u_m=u_0 + \tilde u$. Having
linearized equation~(\ref{eq:G-L_modified}) and assuming that
${\tilde u_r(x,t)= \hat u_r(k)\exp(\Lambda t + ikx)}$, ${\tilde
u_i(x,t)=\hat u_i(k)\exp(\Lambda t+ ikx)}$ we obtain the dispersion
relation
\begin{equation}
\left|
\begin{array}{cc}
\begin{array}{l}
1-u_i^2-3u_r^2-\\2\beta u_iu_r-k^2-\Lambda
\end{array}&
\begin{array}{r}
-(\beta u_r^2+3\beta u_i^2+\\2u_ru_i-\alpha k^2)
\end{array}\\
\\
\begin{array}{l}
\beta u_i^2-2u_iu_r+\\
3\beta u_r^2-\alpha k^2
\end{array}
& \begin{array}{r} 2\beta u_ru_i-u_r^2-\\3u_i^2+1-k^2-\Lambda
\end{array}
\end{array}
\right|=0.\label{Eq:M}
\end{equation}
determining the stability of the homogenous stationary state $u_0$.
The homogenous stationary state $u_0$ is stable if the condition
$\mathrm{Re}\,{\Lambda(k)<0,~\forall k}$ is satisfied.

The evolution of $\mathrm{Re}\,\Lambda(k)$ with the decrease of
$D$-value for $\alpha=1$ and $\alpha=2$ is shown in
Fig.~\ref{fgr:DispRelation},\,\textit{a} and
Fig.~\ref{fgr:DispRelation},\,\textit{b}, respectively. One can
see, that for $\alpha=1$ the homogenous stationary state $u_0$
loses its stability when $D\approx1.5$. In this case the spatial
perturbation with the wave number $k=0$ starts growing
exponentially. As a result, the stationary state $u_0$ becomes
unstable, the spatio-temporal chaos taking place in
system~(\ref{eq:G-L_modified}). \aeh{With the largest Lyapunov
exponent becoming positive} both in the
modified~(\ref{eq:G-L_modified}) and original~(\ref{eq:G-L_noise})
Ginzburg-Landau equation, the noise-induced synchronization regime
in Eq. (\ref{eq:G-L_noise}) is destroyed.

\begin{figure}[tb]
%\vspace*{0.3cm}
\centerline{\includegraphics*[scale=0.35]{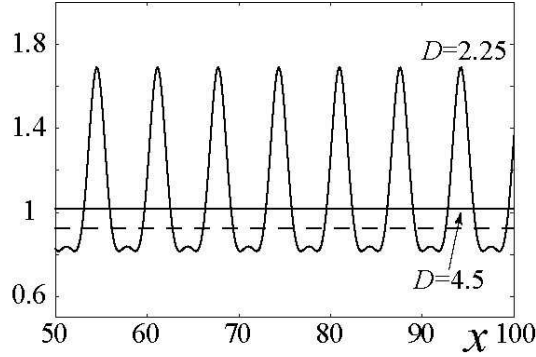}} \caption{The
profiles of the spatial stationary states $|u_0|^2$ and $|u_k(x)|^2$
observed in~(\ref{eq:G-L_modified}) for the different values of the
noise intensity $D$ and $\alpha=2$. Dashed lines correspond to the
unstable state $|z_0|^2$ }\label{fgr:states}
\end{figure}

For the value of the control parameter $\alpha=2$ \alkor{the
homogenous stationary state $u_0$ loses its stability for
$D\approx2.5$ and} the spatial mode with the wave number $k=\pm
0.5$ becomes unstable in contrast to the case of $\alpha=1$
considered before (see Fig.~\ref{fgr:DispRelation},\,\textit{b}).
Therefore, for $\alpha=2$ the periodic spatial state
$u_k(x)=u_k(x+l)$ (where $l$ is close to $2\pi/k$ due to
periodical boundary conditions)\aeh{, which is stationary in time}
replaces the homogenous state $u_0$ in the modified
Ginzburg-Landau equation. The example of the profiles of such
stationary in time but periodic in space states is shown in
Fig.~\ref{fgr:states}. Obviously, for such stationary states the
largest Lyapunov exponent is equal to zero. Evidently, in the
initial Ginzburg-Landau equation driven by noise, $D\zeta(x,t)$,
with the mean value $\langle D\zeta\rangle$ the stationary in time
and periodical in space structure $u_k(x)$ is perturbed by the
fluctuations. Therefore, the spatio-temporal dynamics of $u_k(x)$
looks like aperiodic motion, with the largest Lyapunov exponent
being also equal to zero. Since two identical media, $u(x,t)$,
and, $v(x,t)$, driven by common noise start with different initial
conditions $u(x,0)$ and $v(x,0)$\aeh{, the spatially periodical
structures} do not coincide with each other, i.e., ${u_k(x)\neq
v_k(x)}$, but there is such a shift in space $\delta_0$ depending
on the initial conditions $u(x,0)$ and $v(x,0)$ where
${u_k(x)=v_k(x+\delta_0)}$. Therefore, for ${D_{INIS}<D<D_{NIS}}$
Ginzburg-Landau equations~(\ref{eq:G-L_noise}) driven by common
noise are characterized by zero largest Lyapunov exponent and
their states are not identical. \alkor{If \aeh{the first of the
systems} is shifted along the second one with a certain shift
$\delta_0$ that, depending on initial conditions, satisfies the
requirement $u_k(x)=v_k(x+\delta_0)$, the identical behavior of
both considered systems is observed.}

\alkor{Note, also, a very good agreement between the values of the
noise intensity $D$ corresponding to the loss of the stability of
the homogenous stationary state $u_0$ (see
Fig.~\ref{fgr:DispRelation},\,\textit{b}) and to the point where
the largest Lyapunov exponent becomes equal to zero (see
Fig.~\ref{fgr:Diff_and_CLE_on_D},\,\textit{b}.)}

\alkor{It should be noted that the INIS phenomenon considered
above is determined by the peculiarity of the periodic boundary
conditions. Due to the usage of such a kind of the boundary
conditions, any spatially periodic solution can be moved
arbitrarily along the spatial coordinate, and, therefore, in this
case there \aeh{is additional translational} degree of freedom.
Evidently\aeh{, zero Lyapunov exponent} corresponds to this
translational degree of freedom, whereas all other Lyapunov
exponents are negative. Instantaneous states evolved from the
different initial conditions under the influence of the given
realization of noise can be transformed into each other by means
of the corresponding spatial shift. \aeh{Taking this aspect into
consideration the INIS regime does not seem to be
%From this aspect it follows that the INIS regime can not be
observed in the absence of the translational invariance, i.e., for
the other kind of the boundary conditions.}}

\begin{figure}[tb] %\vspace*{1.0cm}
\centerline{\includegraphics*[scale=0.3]{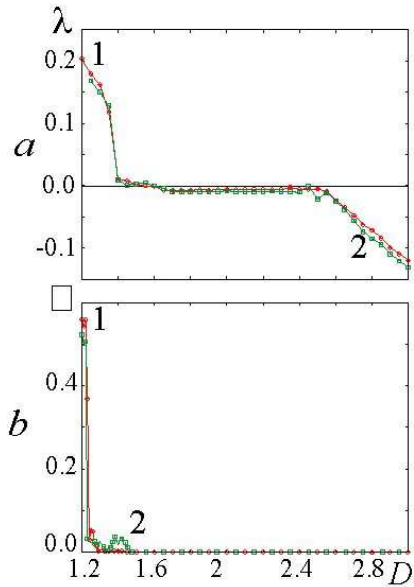}}
\caption{\alkor{(Color online) The dependencies of (\textit{a}) the
largest Lyapounov exponent of the CGLE and (\textit{b}) the averaged
difference (\ref{eq:epsilon}) on the noise intensity $D$ for control
parameter values $\alpha=2$ and $\beta=4$. Curves~1
(\textcolor{red}{$\blacklozenge$}) correspond to the boundary
conditions~(\ref{eq:BoundCondType1}) and curves~2
(\textcolor{green}{$\square$})
--- to the boundary conditions~(\ref{eq:BoundCondType2}),
respectively.}} \label{fgr:LEOtherBoundaryConditions}
\end{figure}

\alkor{To illustrate this statement we have considered the complex
dynamics of Ginsburg--Landau equation~(\ref{eq:G-L_noise}) with the
same set of the control parameter values but with the alternative
types of the boundary conditions eliminating the translational
invariance in the system under study. We have calculated both the
largest Lyapunov exponent and $\lambda$ and the averaged difference
$\varepsilon$ for two types of the boundary conditions:
\begin{equation}
\begin{array}{l}
\displaystyle\left.\frac{\partial u}{\partial
x}\right|_{x=0}=\left.\frac{\partial u}{\partial x}\right|_{x=L}=0\\
\\
\displaystyle\left.\frac{\partial v}{\partial
x}\right|_{x=0}=\left.\frac{\partial v}{\partial x}\right|_{x=L}=0\\
\end{array}
\label{eq:BoundCondType1}
\end{equation}
and
\begin{equation}
\begin{array}{l}
\mathrm{Re}\,u(0,t)=\mathrm{Re}\,u(L,t)=1,\\
\mathrm{Im}\,u(0,t)=\mathrm{Im}\,u(L,t)=1.\\
\mathrm{Re}\,v(0,t)=\mathrm{Re}\,v(L,t)=1,\\
\mathrm{Im}\,v(0,t)=\mathrm{Im}\,v(L,t)=1.\\
\end{array}
\label{eq:BoundCondType2}
\end{equation}
The
obtained curves $\lambda(D)$ and $\varepsilon(D)$ are shown in
Fig.~\ref{fgr:LEOtherBoundaryConditions}.}

\alkor{Evidently, in both cases there is no translation invariance
in the system, and, as a result, the largest Lyapunov exponent is
close to zero, but negative, in the range of $D$-values where for
the periodical boundary
conditions~(\ref{eq:PeriodicBoundaryConditions}) the INIS regime
is observed. The averaged difference $\varepsilon$, in turn, is
equal to zero in this range of the noise intensity values, that
\aeh{shows} the impossibility of the INIS regime existence if
there is no translation invariance in the system. }

\section{The incomplete noise-induced synchronization and the noise characteristics}
\label{sct:NoiseCharacteristics}

To illustrate that the \aeh{onset of} the \alkor{incomplete}
noise-induced synchronization regime is caused by the mean value
of noise only, we examine how the different model noise influence
the considered media described by CGLEs~(\ref{eq:G-L_noise}). One
of the typical probability density is the \aeh{Gaussian} one,
therefore, it seems to be reasonable to consider the dynamics of
the complex Ginzburg-Landau equations~(\ref{eq:G-L_noise}) driven
by common noise with the \aeh{Gaussian} probability distribution
of the real and \aeh{imaginary} parts of the random variable
\begin{equation}
p(\xi)=\frac{1}{\sqrt{2\pi}\sigma}\exp\left(-\frac{(\xi-a)^2}{2\sigma^2}\right).
\label{eq:GaussNoiseDistribution}
\end{equation}
\aeh{Since in Sec.~\ref{sct:GLEs} and Sec.~\ref{sct:Mechanisms} we
were not able to separate the influence of the mean value and
variance from each other, we can do it easily for Gaussian
probability density.}
%
%In contrast with the consideration carried out in
%Sec.~\ref{sct:GLEs} and Sec.~\ref{sct:Mechanisms} where we were
%not able to separate the influence of the mean value and variance
%from each other, for the Gauss probability density it can be done
%easily.
\alkor{To do that, the value of the noise intensity $D=1$
has been fixed. In this case the mean value of the random process
is governed by the choice of $a$-parameter, while the intensity of
noise is determined by the variance $\sigma$.} The random variable
with the required probability density has been generated following
Ref.~\cite{NumericalRecipes:1997}.

\begin{figure}[tb]
%\vspace*{5cm}
\centerline{\includegraphics*[scale=0.3]{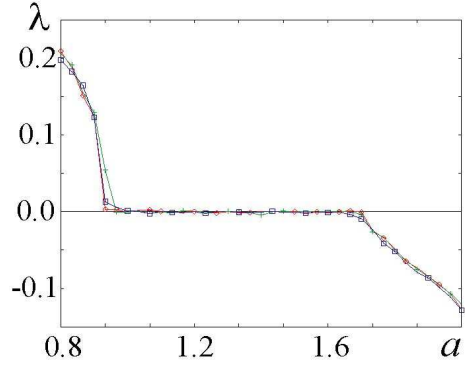}}
\caption{(Color online) The dependencies of the largest Lyapunov
exponent $\lambda$ on the value \alkor{$a$} calculated for the
noise with \aeh{Gaussian} distribution
(\ref{eq:GaussNoiseDistribution}) for the different values of the
variance: $\textcolor{red}{\lozenge}$
--- $\sigma=0.1$, $\textcolor{green}{+}$ --- $\sigma=0.2$,
$\textcolor{blue}{\square}$ --- $\sigma=0.5$. The plateau where
$\lambda=0$ is observed. The control parameter values of complex
Ginzburg-Landau equation are $\alpha=2$, $\beta=4$ }
\label{fgr:LEforGaussian}
\end{figure}

\begin{figure}[tb]
%\vspace*{5cm}
\centerline{\includegraphics*[scale=0.3]{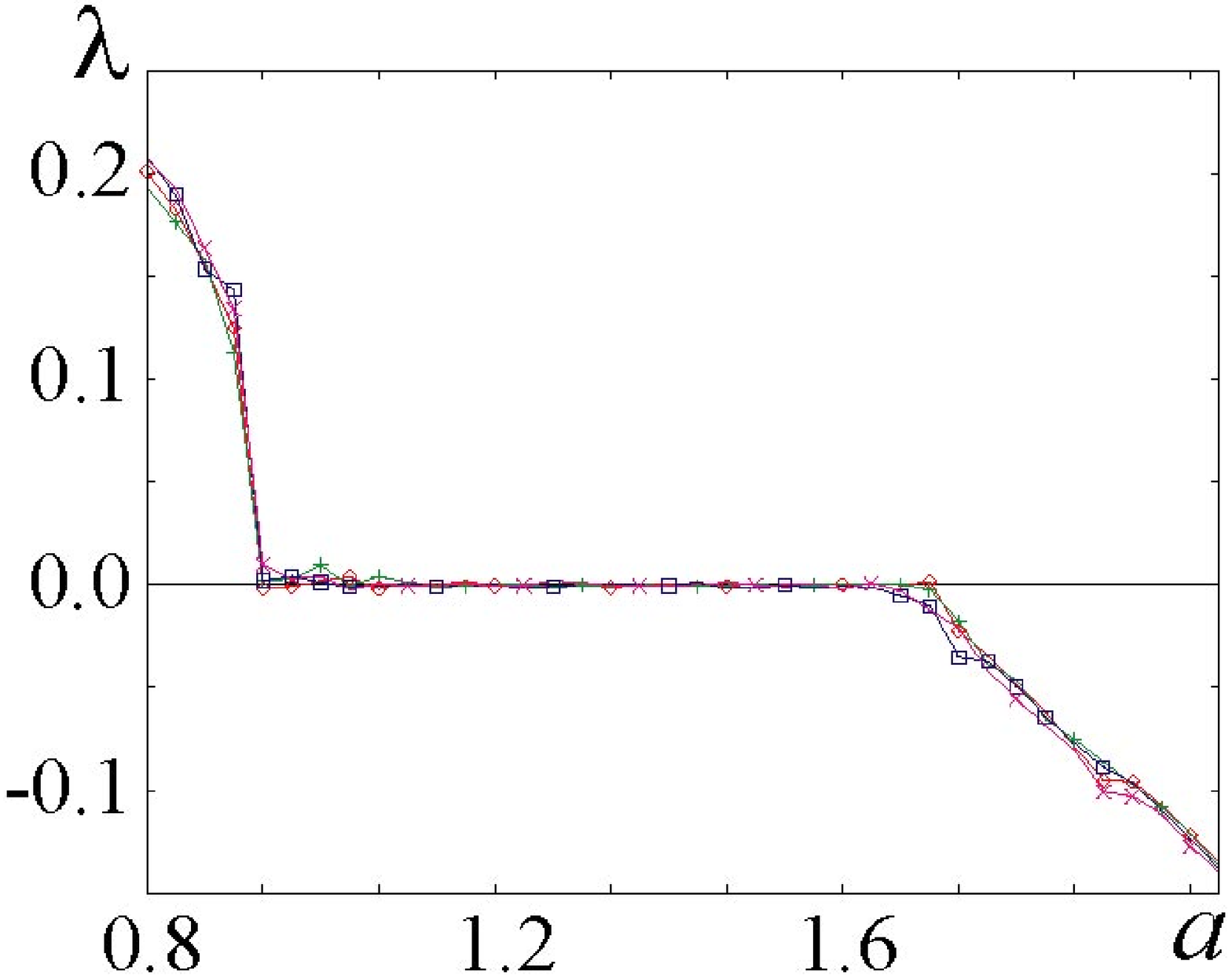}} \caption{(Color
online) The dependencies of the largest Lyapunov exponent $\lambda$
on the value \alkor{$a$} calculated for the uniform distribution of
the random variable with the different values of the variance:
$\textcolor{red}{\lozenge}$
--- $\sigma=0.5$, $\textcolor{green}{+}$ --- $\sigma=1.0$,
$\textcolor{blue}{\square}$ --- $\sigma=1.5$,
$\textcolor{red}{\times}$
--- $\sigma=2.0$. The plateau where $\lambda=0$ is observed. The
control parameter values of complex Ginzburg-Landau equation are
$\alpha=2$, $\beta=4$ } \label{fgr:LEforUniform}
\end{figure}

We have calculated the dependence of the largest Lyapunov exponent
$\lambda$ on the mean value $a$ of the probability
density~(\ref{eq:GaussNoiseDistribution}) for different values of
the variance $\sigma$. The results of these calculations are given
in Fig.~\ref{fgr:LEforGaussian} where the curves $\lambda$ vs.
\alkor{$a$}-value are shown. As it follows from the numerical
calculations the \alkor{incomplete} noise-induced synchronization
is observed in the range \alkor{$a\in[1.0; 1.7]$} (as well as in
the case considered in Sec.~\ref{sct:GLEs}) for all values of the
variance $\sigma$. Therefore, the value of the variance $\sigma$
\alkor{(i.e., the noise intensity)} does not seem to play the key
role in the occurrence of the INIS regime.

Having examined the spatio-temporal behavior of two complex
Ginzburg-Landau equations~(\ref{eq:G-L_noise}) driven by common
noise with the probability
density~(\ref{eq:GaussNoiseDistribution}) we have found for the
range \alkor{$a\in[1.0;1.7]$} that the states of them do not
coincide with each other therefore there is no noise-induced
synchronization as expected. Nevertheless, there is the spatial
shift $\delta$ for which the dynamics of system $u(x,t)$ and
$v(x+\delta,t)$ is identical, which is also the evidence of the
presence of the \alkor{incomplete} noise-induced synchronization
regime.

\alkor{To compare the results described in
Sec.~\ref{sct:Mechanisms} and Sec.~\ref{sct:NoiseCharacteristics}
with each other we have to take into account that the value $2D/3$
has been substituted in Sec.~\ref{sct:Mechanisms} for the mean
value of noise (see the explanation given below
Eq.~(\ref{eq:G-L_modified})). Obviously, if we substitute the same
value $2D/3$ for the mean value $a$ (i.e., denoting $a=2D/3$) in
the case of noise with the \aeh{Gaussian} probability
distribution, we obtain, that the range of the INIS-regime is
${D\in[1.5;2.5]}$ that agrees very well with the results of the
analytical consideration given in Sec.~\ref{sct:Mechanisms}.}

The very same results have been obtained for the uniform
distribution of the random variables with \aeh{the} mean value
\alkor{$a$} and variance $\sigma$ (Fig.~\ref{fgr:LEforUniform}):
there is the range of the values of \alkor{$a$}-parameter where
the largest Lyapunov exponent is equal to zero and the
\alkor{incomplete} noise-induced synchronization regime takes
place.

Based on this consideration we come to conclusion that the
occurrence of the \alkor{incomplete} noise-induced synchronization
regime is determined by the mean value of noise, whereas the
variation of it practically does not play any role.

\section{Conclusion}
\label{sct:Conclusion}

In conclusion, we have reported for the first time a new type of
noise-induced synchronous behavior occurring in the spatially
extended systems. Such a type of \alkor{incomplete} noise-induced
synchronization differs remarkably from all other types of
synchronous behavior known so far. It may be observed in a certain
range of the noise intensity values, where the largest Lyapunov
exponent is equal to zero and the states of two identical
spatially extended systems driven by common noise are different,
although there is an indication of the synchronism: if one of the
systems is shifted along the second one on the certain shift the
identical behavior of the considered systems is observed. The
theoretical equations allowing to explain the mechanism resulting
in such a type of behavior have also been given, and they are in
perfect agreement with the numerically obtained data. The
influence of a different model noise on the occurrence of the
noise-induced synchronization regime is considered. Though the
INIS regime has been  observed here in the complex Ginzburg-Landau
equations driven by common noise with non-zero mean value, we
expect that the very same type of behavior can be observed in many
other relevant circumstances. Since the noise influence may result
in a pattern formation (see,
e.g.,~\cite{Garcia-Ojalvo:1993_Noise}) we suppose that
\alkor{incomplete} noise-induced synchronization can be also
observed for noise with the zero mean value, with the other types
of the spatio-temporal patterns (e.g., traveling waves) being
observed.

\section*{Acknowledgement}

\aeh{We thank Dr. Svetlana V. Eremina for English language
support.} \alkor{We thank the referees of our paper for the useful
comments.} This work has been partially supported by U.S.~Civilian
Research \& Development Foundation for the Independent States of
the Former Soviet Union (CRDF, grant {REC--006}), the Supporting
program of leading Russian scientific schools (project
NSh-355.2008.2), President Program (Grant No. MD-1884.2007.2.) and
Russian Foundation of Basic Research (grant no. 07--02--00044). We
also thank ``Dynasty'' Foundation.

%\bibliography{Export}

\end{document}